\documentstyle[floats,prb,aps]{revtex}

\input epsf
\tighten
\let \mathcal \cal
\begin{document}
\draft
\twocolumn[
\title{Superconducting Upper Critical Field near a 2D van Hove singularity}
\author{R. G. Dias and J. M. Wheatley}
\address{Interdisciplinary Research Centre in Superconductivity, \\
         University of Cambridge, \\
         Madingley Road Cambridge, CB3 OHE, United Kingdom.
}
\date{\today}
\maketitle
\widetext
\begin{abstract}
     \leftskip 54.8pt
     \rightskip 54.8pt
     The superconducting upper critical
     field $H_{c2}(T)$ of a two dimensional BCS
     superconductor
     is calculated in the vicinity of a van-Hove singularity.
     The zero temperature upper critical field
     is strongly enhanced at weak coupling when the Fermi contour
     coincides with van-Hove points, scaling
     as $H_{c2}(0) \propto T_c^{\sqrt{2}}$
     compared to the usual result $H_{c2}(0) \propto T_c^{2}$. The result
     can be interpreted in terms of the
     non-Fermi liquid decay of normal state pair
     correlations in the vicinity of a van-Hove point.

\end{abstract}
\pacs{
\leftskip 54.8pt
\rightskip 54.8pt
PACS numbers: 72.10.-d 74.65.+n 05.40.+j}
]
\narrowtext

The possibility of a strong enhancement of the
superconducting transition
temperature in the vicinity of
van-Hove singularities (vHs)  of
electronic systems has been raised on several occasions.\cite{labbe1,hir}
At a vHs, the Fermi velocity vanishes at points on the Fermi contour
and in 2D, the density of states diverges logarithmically.
While it is tempting to apply this simple model of $T_c$ enhancement
to realizations of quasi-2D systems such
as Copper-Oxide superconductors,\cite{labbe2,tsu} it
is often pointed out
that three-dimensionality and scattering effects have the effect of
smearing out the vHs. Also, the electron kinetic energy favors low
density of states at the Fermi level
and structural distortions may be expected
to occur in order to push the vHs
below the Fermi level.\cite{fried}

Despite these limitations, the 2D vHs
provides a simple and well-understood
example of a normal Fermi system which
possesses anomalous normal state correlations and
therefore deserves to be understood in detail.
The origin of non-Fermi liquid
effects in this case is the absence of a well-defined Fermi velocity
scale, which is in turn due to the presence
of an underlying lattice and the absence of
Galilean invariance.
It has recently been shown that non-Fermi liquid
behavior in the normal state leads to deviations
from the usual quasi-parabolic
mean-field $H_{c2}(T)$ curve of
Fermi liquid BCS (FL-BCS) superconductors
found by Gorkov.\cite{whea,sudbo}
Interestingly, deviations from the parabolic
shape have been observed in layered superconductors,\cite{expts2}
including both Copper-Oxides\cite{expts} and
$k\mbox{-(BEDT-TTF)}_2\mbox{Cu(NCS)}_2$.\cite{expts1}
As emphasized previously,\cite{whea} the mean field
upper critical field is a normal state property which
probes both the spatial (via the magnetic
length) and energy (temperature)
dependence of pair correlations. It may therefore prove
an effective probe of non-Fermi liquid normal effects.

In the present paper we show that anomalous correlations
in the vicinity of a 2D vHs lead to spectacular deviations from the
FL-BCS result.
Following the semi-classical Gorkov approach,\cite{gorkov,gorkov2}
the superconducting transition corresponds to
the appearance of non-trivial solutions of the equation
\begin{equation}
         \Delta({\bbox r}) = g\int d^2 {\bbox r^\prime}
         K_\beta({\bbox r^\prime})
         e^{i [{\bbox A }({\bbox r })+{\bbox A }({\bbox r^\prime })]
         ({\bbox r^\prime}-{\bbox r })}
         \Delta( {\bbox r} + {\bbox r^\prime})
         \label{eq:gorkov1}
\end{equation}
where $ \beta $ is the inverse temperature and
$g$ the pairing interaction. The magnetic field
effect is included in a semi-classical form as a
phase term\cite{gauge}
$ \exp({i [{\bbox A }({\bbox r })+ {\bbox A }({\bbox r^\prime })]
({\bbox r^\prime}-{\bbox r })})$. This approximation neglects
Landau level quantitization which is expected to become
important in the high field limit. The latter behavior has not yet
been observed experimentally within the range of
fields presently accessible and the semi-classical
approximation will be used in this paper.
$K_\beta(r)$ is the {\it normal state} pair
propagator in the absence of an external magnetic
field:
\begin{equation}
          K_\beta({\bbox r}) = \beta^{-1} \sum_{\omega_n}
          g({\bbox r},\omega_n) g({\bbox r},-\omega_n).
\end{equation}
Here $g(r,\omega_n)$ is the real space single
particle green function at Matsubara frequency
$\omega_n$.
The highest eigenvalue of Eq.~(\ref{eq:gorkov1}) determines the
upper critical field. We work in the Landau
gauge ${\bbox A }=(0,Hx,0)$. Making use of
the degeneracy of the gap function,\cite{gorkov} one can write
\begin{equation}
         \widetilde{\Delta}(x) = g\int dx^\prime
         \widetilde{K}_\beta[x^\prime-x,-H(x+x^\prime)]
         \widetilde{\Delta}(x^\prime),
         \label{eq:gorkov2}
\end{equation}
where $\widetilde{\Delta}(x)$ is the gap function integrated over $y$
and  $\widetilde{K}_\beta (x,k_y)$ is the
Fourier transform of $K_\beta (x,y)$. The latter can be
expressed as
\begin{eqnarray}
        \widetilde{K}_\beta(x,k_y) & = & 2 \int d\omega \int dp_y
        \tanh(\beta \omega/2) \times \nonumber\\
        & & A(x,p_y+k_y,\omega) B(x,-p_y,-\omega),
        \label{eq:fouAB}
\end{eqnarray}
where $A(x,k_y,\omega) = {1 \over 2 \pi} \int dk_x \exp(i k_x x)
{\hbox {Im}} g({\bbox k},\omega)$
and $B(x,k_y,\omega) = {1 \over 2 \pi} \int dk_x \exp(i k_x x)
{\hbox {Re}} g({\bbox k},\omega)$.

When the
Fermi energy $E_F$ is close enough to the van-Hove singularity
$E^{vh}$, one can express
$E({\bbox k})-E^{vh} = q_x q_y$ where
${\bbox q}={\bbox k}-{\bbox k^{vh}}$ and
the values of the momentum are restricted to a small
region around the saddle point ${\bbox k^{vh}}$
by a cutoff.
For simplicity, we assume that $2{\bbox k^{vh}}$
is a reciprocal lattice vector, but the results
are independent of this assumption.
We now have
\begin{eqnarray}
        A(x,q_y,\omega) &=& -{ 1 \over 2 \vert q_y \vert}
        e^{i {x (\omega + \delta) \over q_y} }
        e^{i {\bbox k}_{x}^{vh} x}, \nonumber\\
        B(x,q_y,\omega) &=& i {\hbox {sgn} }
        ({x \over q_y}) A(x,q_y,\omega),
\end{eqnarray}
with $\delta=E_F -E^{vh}$.
The pair propagator  is
\begin{eqnarray}
        \widetilde{K}_\beta(x,2q_y) & = &  {1 \over 2}
        \int_{\vert q_y \vert }^{p_{o}} dp
        {2 \pi /\beta \over \left( p^2 - q_y^2 \right)}
        { \cos \left[ \delta \vert x \vert {p \over
        p^2 - q_y^2 } \right]
        \over \sinh \left[ {2 \pi \over \beta} \vert x \vert {p \over
        p^2 - q_y^2 } \right] }.
        \label{eq:kernel}
\end{eqnarray}
We have introduced a cutoff $p_o$ for the $y$ component
of the momentum.
Note that doping away from the vHs introduces a
oscillating term in the integral and $K_\beta (x,k_y)$ decreases.
The effects of the temperature and the doping
in the integral are similar as they provide a cutoff
for large $x$.
One expects $H_{c2}(T)$ to be rather
insensitive to $T$ for $T<\delta$, that is,
$H_{c2}$ remains approximately constant to a
temperature of the order of $\delta$.
For $T>\delta$, we have the opposite, that is,
the $H_{c2}$ curves are fairly independent of the
filling. At zero temperature the asymptotic
decay of Eq.~(\ref{eq:kernel}) in real space is
$K \sim \vert x \vert^{-1} \vert y \vert^{-1}$.
This is a slow decay relative to the
 2D Fermi Liquid
(circular Fermi surface) result $K \sim (x^2 + y^2)^{-1}$.
\begin{figure}[bt]
       \begin{center}
       \leavevmode
       \hbox{%
       \epsfxsize 2.5in \epsfysize 2.0in \epsfbox{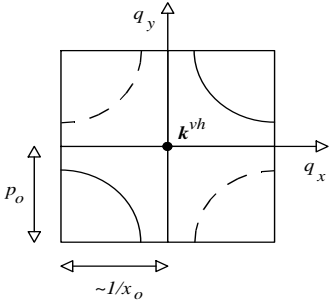}}
       \end{center}
       \vspace{-.4cm}
       \caption{Constant energy contours of the
       quasiparticle dispersion near
       a 2D van Hove point ${\protect\bbox k^{vh}}$.
       $E({\protect\bbox k})-E^{vh} = q_x q_y$.
       The values of ${\protect\bbox q}$ are restricted to a region around
       the saddle point
       by the cutoffs $p_o$ and $x_o$.}
       \label{fig:one}
\end{figure}

Firstly, we evaluate the critical
temperature at zero field when the vHs lies at the
Fermi level. We need the ${\bbox q} = 0$ pair
propagator (or pair susceptibility):\cite{uniprop}
\begin{eqnarray}
        {\mathcal K}_\beta(0,0) & = &  {\int_{x_0}^{\infty} dx
        \widetilde{K}_\beta(x, q_y=0) }\nonumber\\
        & = & {\int_{x_o}^{ \infty}
        {d x \over x}
        \log \left[  \tanh \left( {x\pi \over { \beta_c p_o}}
        \right) \right]^{-1}}. \nonumber
\end{eqnarray}
Here we have introduced a short distance cutoff
$x_0$ in the x-direction.\cite{poxo}
At temperatures low compared with the bandwidth
$x_o { \pi / { \beta p_o}} \ll 1$, the main contribution
to the integral results from the small $x$ region and
$ {\mathcal K}_\beta(0,0) \sim 1 / 2 \log^2 \left[
{x_o \pi} / {\beta p_o} \right] $.
The zero field critical temperature is determined by the
relation $1=g {\mathcal K}_{\beta_c}(0,0)$ and
one recovers the standard result 
$T_c = \pi^{-1} x_o^{-1} p_o \exp(-\sqrt{2/g})$ compared to the
weak coupling FL-BCS result $T_c \propto \exp(-1/\widetilde{g})$,
where $\widetilde{g}\propto g$.

The pair propagator [Eq.~(\ref{eq:kernel})] can be evaluated
in closed form at zero temperature and zero $\delta$.
In this case, the gap equation [Eq.~(\ref{eq:gorkov2})]
simplifies to
\begin{equation}
         \widetilde{\Delta}(x) = g
         \int dx^\prime {1 \over 2 \vert x^\prime-x \vert}
         \log \left[{2 p_o \over H \vert x^\prime+x \vert}\right]
         \widetilde{\Delta}(x^\prime),
         \label{eq:gapzeroT}
\end{equation}
with $\vert x^\prime-x \vert>x_o$.
A lower bound for $H_{co}$ can be obtained by the variational method.
A physically reasonable form for the gap function is
$\widetilde{\Delta}(x)= {1/\sqrt{2 a}} \Theta(a - \vert x \vert)$,
where $a$ is a length to be determined. The maximum
eigenvalue is obtained for $a=e\sqrt{x_o p_o/2H}$ which leads to
${1 / g} \gtrsim {1 / 4} \log^2\left({e^2 p_o / 2 x_o H_{co}}\right).$
Thus
$H_{co}\sim (e^2 p_o/2 x_o) \exp(-2 g^{-{1\over 2}})$. This implies
\begin{equation}
{x_o H_{co} \over p_o} \sim \left({x_o T_{co} \over p_o}\right)^{\sqrt{2}}.
\label{eq:zeroT}
\end{equation}
\begin{figure}[bt]
       \begin{center}
       \leavevmode
       \hbox{%
       \epsfxsize 3.0in \epsfysize 2.5in \epsfbox{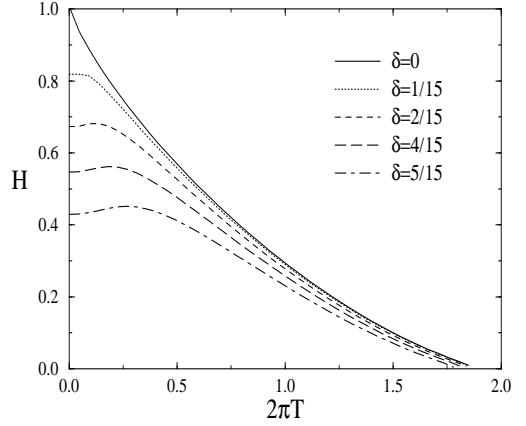}}
       \end{center}
       \caption{Variation of the $H_{c2}$ curve as
       a function of $\delta$, for $p_o/x_0=20$ and
       moderate coupling $g=0.2$, obtained by
       numerical lanczos solution of
       Eq. (3). Notice that
       $H_{co}$ has a much stronger dependence on
       $\delta$ that $T_{co}$. Upward curvature is
       observed in all curves down to a temperature
       of the order of $\delta$. For higher $\delta$,
       there is a small increase of $H_{c2}$ with
       temperature as the temperature averaged density
       of states increases.}
       \label{fig:two}
\end{figure}
This result should be contrasted with the usual
FL-BCS scaling $H_{co} \sim T_{co}^2$. By
expanding the kernel [Eq.~(\ref{eq:kernel})]
around the zero temperature critical point,
one can obtain the low temperature behavior
of $H_{c2}$ and behaves as $H_c-H_{co}\sim
-T \log{(H_{co}/T)} / \log{(x_o H_{co}/p_o)}$.
$H_c(T)$ {\it does not saturate} as $T$ goes to zero.

For temperatures close to the zero field critical
temperature $T_{co}$, one can follow the Gorkov microscopic
derivation of the Ginzburg-Landau equation to obtain
\begin{equation}
         a \widetilde{\Delta}(x)+
         b (\partial^2_x -H^2x^2)
         \widetilde{\Delta}(x) = 0,
\end{equation}
where
\begin{eqnarray}
         a &=& (T-T_{co}) {d \over dT} \left[
         \int d^2 {\bbox r} K_\beta({\bbox r})
         \right]_{T_{co}}, \nonumber \\
         b &=& {1 \over 2} \int d^2 {\bbox r} K_\beta({\bbox r})
         x^2 \cong
         {1 \over 2} \int d^2 {\bbox r} K_\beta({\bbox r})
         y^2 . \nonumber
\end{eqnarray}
Here, we have used the fact that $K_\beta({\bbox r})$
is symmetric in the transformation $x \rightarrow y$.
Setting $p_o\sim{1\over x_o}$,
then $b_x \sim b_y$.
For our van-Hove model with $\delta=0$, we have
\begin{eqnarray}
         a &\sim & {T_{co}-T \over T_{co}} \log
         \left[ {p_o \over \pi x_o T_{co}} \right] , \nonumber \\
         b &\sim & \left( { p_o\over 2\pi T_{co}} \right)^2 , \nonumber \\
         {dH_c \over dT} &\sim& - {x_o T_{co}\over p_o}
          \log \left[ {p_o \over \pi x_o T_{co}} \right] .
          \label{eq:slope}
\end{eqnarray}
Note that for the 2D isotropic BCS superconductor,
we would have $a \sim \rho(E_F) (T_c-T)/T_c$. The vHs
at the Fermi level is averaged out at a
finite temperature $T_{co}$ and so, in the previous expressions,
the density of states is replaced by an effective
density of states $\log [\pi {x_o T_{co} / p_o}] $
giving a weak enhancement
of the slope.
Eqs.~(\ref{eq:zeroT}) and (\ref{eq:slope}) show that there is an
anomalous enhancement and consequently
upward curvature of the upper critical field and that
this upward curvature is strongly enhanced at weak
coupling (low $T_c$).
Thus in contrast to FL-BCS, {\it normalized plots of
${H \over H_{co}}\left({T \over T_{co}}\right)$
do not fall on to a universal curve} for
different couplings.

To extend the above analysis to all temperatures and confirm the
variational analysis requires a
numerical approach.
Discretizing in the $x$ direction, $x_n=n\epsilon$
where $n$ are integers, the integral gap equation
becomes
\begin{equation}
         \widetilde{\Delta}_i = g\sum_{ij}
         \widetilde{K}_{ij}
         \widetilde{\Delta}_j.
         \label{eq:numerical}
\end{equation}
While the system is now discrete it remains infinite
dimensional.
However, the system can be truncated for finite $H$ because we
need only consider solutions
$\widetilde{\Delta}_i$ which are localized
about the origin on a scale of the magnetic length.

The highest eigenvalue and corresponding
gap function of Eq.~(\ref{eq:numerical})
were obtained by the
Lanczos method.
\begin{figure}[bt]
       \begin{center}
       \leavevmode
       \hbox{%
       \epsfxsize 3.0in \epsfysize 2.5in \epsfbox{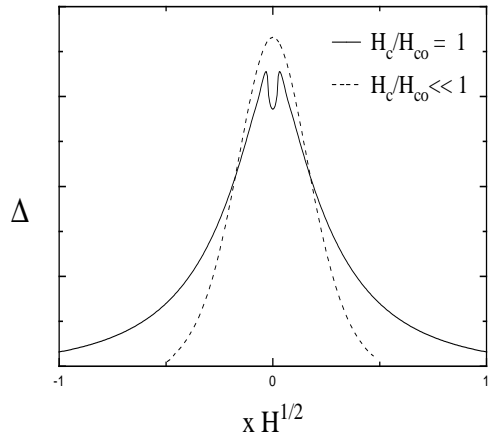}}
       \end{center}
       \caption{Gap functions $\widetilde{\Delta}(x)$ for
       two points of the $H_{c2}$ curve
       with $p_o/x_0=20$, $g=.2$ and $\delta=0$.
      One of the points is near the top of the
      $H_{c2}$ curve and the other near the bottom.
       Although the range of the gap
       functions is roughly the same (of the
       order of the magnetic length in each case),
       the functional forms of the two curves
       are different. For low $H_c/H_{co}$
       we have the conventional behavior
       $\widetilde{\Delta}(x) \sim e^{-  x^2H}$, while for
       $H_c/H_{co} \sim 1$, the gap function has an anomalously slow
       decay.}
       \label{fig:three}
\end{figure}
The results are displayed
in Figs.~\ref{fig:two} and \ref{fig:three}. In
Fig.~\ref{fig:two}, we have $H_{c2}$
curves for different values of $\delta$ and fixed
coupling $g$. $H_{c2}$ saturates when $T<\delta$.
When $\delta=0$, that is, when the Fermi energy
coincides with the vHs, there is  no saturation
and the $H_{c2}$ curve shows an upward curvature
through the complete temperature range in complete agreement
with the variational calculation.

A qualitative understanding of these results
is obtained as follows. In FL-BCS, the upper critical
field is given by
$1=g \rho(E_F) \log (H+T^2)$.
In the vHs case, $\rho(E)$ is logarithmic divergent.
Finite magnetic field or temperature
smear out the density of states and $\rho(E_F)$ should be replaced
by $\log (H+T)$ at zero $\delta$.
Thus $1\sim g \log (H+T) \log (H+T^2)$.
This leads to $H_{co} \sim T_{co}^{\sqrt{2}}$ as found above.

In conclusion, we have studied the suppression of
the mean field superconducting instability
of a clean weak coupling BCS
superconductor in finite magnetic field at a 2D vHs. The
upper critical field is strongly enhanced
relative to $T_c$; this enhancement is described by
the relation $H_{c2}(0) \sim T_c^{\sqrt{2}}$.
The upper critical field falls linearly with temperature near $T=0$.
These effects disappear rapidly when the
Fermi level is tuned away from the Fermi level; more
precisely, they are absent when $T_c < \delta$ defined above.
The 2D vHs provides a simple example of a
system where non-Fermi liquid normal state
correlations show up strongly in the temperature
dependence of the superconducting upper critical field.

RD would like to thank
Junta Nacional de Investiga\c{c}\~{a}o Cient\'{\i}fica e Tecnol\'{o}gica
(Lisbon) for financial support.


\begin{references}
\bibitem{gauge} There is some freedom in the choice
                of the phase term. An equivalent expression
                would be $ \exp\{i [2 {A }({ r })]
                ({r^\prime}-{r })\}$ and it would
                lead to the same results.
\bibitem{uniprop} An equivalent expression is
          $$ {\mathcal K}(0,0) = {\int_{0}^{ p_o}
          dp_y {1 \over p_y} \log \left[
          { \tanh \left( {\pi \over \beta}
          {x_o \over p_y} \right) } \right]^{-1} }, $$
          where the integrand is the 2D isotropic uniform pair propagator
          with Fermi velocity $v_F=p_y$.
\bibitem{poxo} It can be shown that $H_{c2}(T)$
               depends only on the ratio $p_o/x_o$.
\end{references}
\end{document}